\documentclass[aps,pra,notitlepage,twocolumn]{revtex4-1}
\usepackage{amsmath,amssymb,graphicx}

\begin{document}

\newcommand{\ket}[1]{| #1\rangle}
\newcommand{\modsq}[1]{\left| #1 \right|^2}
\newcommand{\av}[1]{\langle #1 \rangle}
\newcommand{\pd}{{\phantom{\dagger}}}
\newcommand{\ketbra}[2]{|#1\rangle\!\langle#2|}

\title{Sensitivity bounds of a spatial Bloch-oscillations Atom Interferometer}

\author{I. Na{\l}\c{e}cz$^1$, L. Masi$^2$, G. Ferioli$^2$, T. Petrucciani$^2$, M. Fattori$^2$ and J. Chwede\'nczuk$^1$}
\affiliation{$^1$Faculty of Physics, University of Warsaw, ul. Pasteura 5, PL--02--093 Warszawa, Poland\\
  $^2$LENS and Dipartimento di Fisica e Astronomia, Universit\'a di Firenze, 50019 Sesto Fiorentino, Italy  }

\begin{abstract}
  We study the ultimate bounds on the sensitivity of a Bloch-oscillation atom interferometer where the external force is estimated from the measurement of the on-site atomic density.
  For external forces such that the energy difference between lattice sites is smaller than the tunneling energy, the atomic wave-function spreads over many lattice sites, 
  increasing the separation between the occupied modes of the lattice and naturally enhancing the sensitivity of the interferometer. 
  To investigate the applicability of this scheme we estimate the effect of uncontrolled fluctuations of the tunneling energy and the finite resolution of the atom detection. 
  Our analysis shows that a horizontal lattice combined with a weak external force allow for high sensitivities. Therefore, this setup is a promising solution for 
  compact devices or for measurements with high spatial resolution. 
\end{abstract}

\maketitle

\section{Introduction}
Atom interferometry is a powerful tool for sensing of gravity, inertial forces and electro-magnetic fields~\cite{Peters_Nature, Gustavson, Treutlein, Hardman}, or measuring 
the fundamental constants~\cite{Tino_nature, fine5} and testing the foundations of physics~\cite{Hamilton, Rasel, Zhan}. Free-falling atom interferometers offer the highest sensitivity and 
are the core technology in many experiments aiming at accurate gravimetry~\cite{Bouyerthree}, gradiometry~\cite{Tino, Pereira, Rosi_2017}, measurements of rotations~\cite{Landragin}, 
inertial navigation~\cite{Bouyer} gravitational wave detection\cite{Bouyertwo}, general relativity tests~\cite{Rasel2, Loriani} and geodesy from space missions~\cite{Raseltwo, refId0}. 
However their sensitivity scales with the size of the interrogation area and this limits their use in application where high spatial resolution is required. Trapped atom interferometers 
are a valuable alternative~\cite{Cronin}. Different schemes have been implemented including Bloch oscillations~\cite{Ferrari}, double well traps~\cite{PhysRevLett.105.243003, berrada2013integrated, Kim2017} 
and Wannier Stark atom interferometers~\cite{Pelle, Ivanov}. 

Although arbitrarily long interrogation times can lead to high sensitivity, these schemes have so far suffered from some limitations, like decoherence induced by interactions~\cite{Javanainen}, 
trapping potential imperfections~\cite{Shin} and limited separations between the spatial modes of the interferometer~\cite{Cronin}. Solutions to this last problem have been addressed in several proposals 
and investigated in many current experiments. All these methods require combinations of optical lattices~\cite{Bresson, PhysRevA.94.043608}, harmonic traps~\cite{Weidong} or in general dynamically varying 
trapping potentials with high quality and stability~\cite{Spagnolli}. It is desirable to develop a scheme where a single optical lattice is used, since it reduces the experimental requirements 
on a trapping potential and because the high control of the lattice frequency naturally increases the accuracy of the measurements. 
Bloch-oscillations atom interferometry, where the periodic oscillations of the momentum distribution of the atoms is observed, fulfill such requirement since only a lattice, 
plus the external force to be measured, is needed to operate the sensor~\cite{dahan1996bloch}. As demonstrated in a recent paper~\cite{JanChw}, 
the sensitivity depends only on the initial coherence length $\xi$ of the source. However its scaling with the initial temperature $T$ of the gas (i.e. $\xi = h/\sqrt{2\pi m k_B T}$ where $m$ is the mass 
of a single atom, $k_B$ is the Boltzman constant and $h$ is the Planck constant), make unrealistic any significant improvement of Bloch-oscillation interferometry beyond the state of the art.

Triggered by recent works~\cite{Preiss, Geiger}, where two groups have reported the observation of the spatial evolution of the gas in-trap, we investigate
the ultimate bounds on the sensitivity of a spatial Bloch-oscillation interferometer (SBOI) where we detect the on-site atomic density rather than the atomic momentum distribution. 
In the case of horizontal lattice operation, for weak external forces, i.e., such that the energy difference between lattice sites is smaller than the tunneling energy, the atomic 
wave-function spreads over many lattice sites, naturally increasing the separations between the occupied modes of the lattice. Our analysis shows that this evolution, together with the capability to
address single 
sites, leads to high sensitivities, making the scheme we propose a promising solution for compact devices or for detection of weak forces with high spatial resolution.

The paper is organized as follows. In Section~\ref{sec.mod} we present the main results of this work. In particular in Section~\ref{sec.ham} we introduce the Hamiltonian 
and characterize the evolution of the system. In Section~\ref{sec.est} we derive the ultimate bound of the sensitivity (Section~\ref{sec.qfi})
and compare  it to an estimation protocol based on the counting of atoms in each site of the lattice (Section~\ref{sec.numb}) or on the measurement of the width of the atomic cloud (Section~\ref{sec.width}). 
In Section~\ref{sec.init} we study how the sensitivity depends on the initial distribution of atoms in the lattice and on the tunneling energy between the sites 
(Section~\ref{sec.analytic}). In Section~\ref{sec.exp} we investigate the most favorable experimental configuration (Section~\ref{sec.configuration}), the effect of a fluctuating tunneling energy (Section~\ref{sec.tun}) and of a non-ideal atom counting (Section~\ref{sec.res}), the dependence of the sensitivity on the lattice spacing (Section~\ref{sec.lat_spec}) and a configuration of optimal performance (Section~\ref{sec.example}). Finally we summarize our findings and conclude our analysis in Section~\ref{sec.ack}. Some details of calculations, omitted for clarity in the text, are presented in the Appendix.

\section{Model and sensitivity}\label{sec.mod}

\subsection{Hamiltonian}\label{sec.ham}
Our starting point is the Hamiltonian of an ultra-cold Bose gas of $N$ atoms in a one-dimensional optical lattice in presence of an external force $mg$ 
\begin{align}\label{eq.ham0}
  \hat H=\int\!\!\hat\Psi^\dagger(x)\left[-\frac{\hbar^{2}}{2m}\Delta+V_{\rm lat}(x)+mgx\right]\hat\Psi(x)\,dx,
\end{align}
where $V_{\rm lat}(x)$ is the optical lattice potential, $g$ is the acceleration and $\hbar=h/2\pi$ is the reduced Planck constant. In the tight-binding approximation, we represent the field operator
as a series of operators annihilating an atom in the $k$-th site
\begin{align}\label{eq.op}
  \hat{\Psi}(x)=\sum_kw_k(x)\hat a_k,
\end{align}
where $w_k(x)$ is the Wannier-like spatial wave-function localized in the $k$-th well. Here and below we consider the infinite lattice, hence the sum runs from $-\infty$ to $+\infty$. Upon the substitution of Eq.~\eqref{eq.op} into Eq.~\eqref{eq.ham0} we obtain, up to the leading order of the overlap
of the Wannier functions
\begin{align}\label{eq.ham}
  \hat H=-J\sum_k\left[\hat a_k^\dagger \hat a^\pd_{k+1}+\hat a^\pd_k\hat a^\dagger_{k+1}\right]+\delta\sum_kk \hat a_k^\dagger \hat a^\pd_k.
\end{align}
The two coefficients $J$ and $\delta$ correspond to the hopping energy and the energy difference between neighboring sites, respectively, and are equal to
\begin{subequations}
  \begin{align}
    J&=\int\!\! w^*_k(x)\left[-\frac{\hbar^{2}}{2m}\Delta+V_{\rm lat}(x)\right]w_{k+1}^{\phantom{*}}(x)\,dx,\\
    \delta&=mg\int\!\! \Big[|w_k(x)|^2-|w_{k+1}(x)|^2\Big]x\,dx\simeq mgx_0,
  \end{align}
\end{subequations}
where $x_0$ is distance between the adjacent wells. This Hamiltonian~\eqref{eq.ham} sets the dynamics of the Bloch oscillations of the gas,
which we assume to be a pure Bose-Einstein condensate (BEC). The initial state reads
\begin{align}
  \ket{\vec\alpha(0)}=\frac{1}{\sqrt{N!}}\Big[\vec{\alpha}(0)\hat{\vec{a}}^\dag\Big]^{N}\ket{0},
\end{align}
where $\vec\alpha(0)$ is a vector of complex amplitudes ($\modsq{\alpha_k(0)}$ sets an initial density of atoms at site $k$) and $\hat{\vec{a}}^\dag$ is a corresponding vector of creation operators. Furthermore,
$\ket{0}$ denotes the vacuum state.
The solution of the Schr\"odinger equation
\begin{align}\label{eq.sch}
  i\hbar\partial_t\ket{\vec \alpha(t)}=\hat H\ket{\vec\alpha(t)}
\end{align}
takes a particularly simple form
\begin{align}\label{eq.sol}
  \alpha_k(t)=\sum_jU_{kj}(t)\alpha_j(0),
\end{align}
since the Hamiltonian in Eq.~\eqref{eq.ham} is quadratic. Here $U_{kj}(t)$ is the matrix element of the evolution operator
\begin{align}\label{eq.evo}
  \hat U(t)=e^{-i\frac{\hat H t}\hbar}.
\end{align}
We now discuss how the acceleration $g$ can be estimated
from the measurement of the on-site atomic population rather than releasing the BEC from the lattice as it is generally done in ultra-cold atom experiments \cite{dahan1996bloch}.

\subsection{Estimation}\label{sec.est}
In this Section we estimate the theoretical sensitivity of an SBOI. In~\ref{sec.qfi} we exploit the quantum Fisher information (QFI) to  calculate the ultimate bound, optimizing
over all possible measurements and detection protocols~\cite{braunstein1994statistical}. 
In~\ref{sec.numb} and~\ref{sec.width} we estimate the sensitivity provided by a measurement of the populations in each site and by the width of the cloud, respectively. 
Finally, in~\ref{sec.analytic} we discuss the dependence of the sensitivity on the number of initially populated sites.  

\subsubsection{Ultimate sensitivity}\label{sec.qfi}
The highest precision an interferometer can achieve is given by the inverse of the QFI. For pure states, as considered here, it reads
\begin{align}\label{eq.qfi}
  F_q=4\left(\av{\hat h^2}-\av{\hat h}^2\right)\equiv4\Delta^2\hat h,
\end{align}
where the average is calculated at time $t$ using the expression $\ket{\vec\alpha(t)}$ and where $\hat h$ is the generator of the interferometric transformation
set by the evolution operator introduced in Eq.~\eqref{eq.evo}
\begin{align}\label{eq.gen}
  \hat h=i\frac{\partial\hat U(t)}{\partial g}\hat U^\dagger(t).
\end{align}
The calculation of $\hat h$ together with the Cramer-Rao lower bound~\cite{braunstein1994statistical} gives the formula
\begin{align}\label{eq.crlb}
  \frac{\Delta g_{\rm opt}}{g}=\frac1{\sqrt{F_q}}\frac1{g}.
\end{align}
Using this formula we determine the ultimate sensitivity. As a first case we consider a BEC of $N=4\times10^4$ atoms initially localized in one site and take $J=\delta$. 
The result, obtained by numerically solving the Schr\"odinger equation~\eqref{eq.sch}, is drawn with a dotted grey line in Fig.~\ref{fig.pure} in the time interval $t\in[5.5,7.5]T_B$, 
where  $T_B=mgx_0/h=\delta/h$. Note that the ultimate sensitivity monotonously improves with time. This reflects the growth of information about $g$, deposited in the system, and formally 
it is a consequence of the action of the derivative in Eq.~\eqref{eq.gen} of the evolution operator~\eqref{eq.evo}.
We now compare this ultimate bound with the sensitivity calculated with two different measurement schemes.

\subsubsection{Site-resolved atom number measurement}\label{sec.numb}
First we assume to detect, via an in-situ measurement, $n^{(j)}_k$---the number of atoms  in each site. Here, $k$ labels the sites and $j$ indexes the measurements. 
The outcomes are averaged over $\nu$ repetitions, giving
\begin{align}\label{eq.mean.m}
  \overline n_k=\frac1\nu\sum_{j=1}^\nu n^{(j)}_k.
\end{align}
According to the central limit theorem, if $\nu$ is large, the probability for obtaining $\overline n_k$ is Gaussian, 
\begin{align}
  p(\overline n_k)=\frac1{\sqrt{2\pi\Delta^2\hat n_k}}e^{-\frac{(\overline n_k-\av{\hat n_k})^2}{2\Delta^2n_k/\nu}}.
\end{align}
Here $\av{\hat n_k}$ and $\Delta^2\hat n_k$ are true values (i.e., calculated asymptotically at $\nu\rightarrow\infty$) of the on-site atom number mean and fluctuations, respectively. This set of outcomes is used to
construct the likelihood function
\begin{align}
  \mathcal L(\tilde g)=\sum_k\log\left(p(\overline n_k)\right).
\end{align}
The gravitational acceleration, which is to be estimated, is treated as a free parameter $\tilde g$ (it enters through $\av{\hat n_k}$ and $\Delta^2\hat n_k$, while $\overline n_k$'s, 
deduced from the experiment, depend on the true value of $g$). The parameter is estimated as this value of $\tilde g$ (denoted by $\tilde g_{\rm ml}$) at which the likelihood function reaches its maximum. 
It is called the maximum likelihood estimator, it is unbiased and has a sensitivity
\begin{align}\label{eq.sens.ml}
  \Delta^{2}\tilde g_{\rm ml}=\frac1\nu\frac{F_1+F_2}{F_1^2}.
\end{align} 
Here the two components of the sensitivity are
\begin{subequations}\label{eq.f12}
  \begin{align}
    F_1&=\sum_k\frac{(\av{\hat n_k}')^2}{\Delta^2\hat n_k}\label{eq.f1},\\
    F_2&=\sum_{k\neq l}\frac{\av{\hat n_k}'}{\Delta^2\hat n_k}\frac{\av{\hat n_l}'}{\Delta^2\hat n_l}\sigma_{k,l}^{2}\label{eq.f2},
  \end{align}
\end{subequations}
where the prime denotes the derivative over the parameter, i.e., $\av{\hat n_k}'=\frac{\partial}{\partial g}\av{\hat n_k}$ and $\sigma_{k,l}^{2}=\av{\hat n_k \hat n_l}-\av{\hat n_k}\av{\hat n_l}$ 
is the cross-correlation of the site occupations~\cite{chwed_njp}. 

The moments of the atom number operator $\hat n_k=\hat a^\dagger_k\hat a_k^\pd$ that enter Eqs~\eqref{eq.f12} read
\begin{subequations}\label{eq.coeffs}
  \begin{align}
    &\av{\hat n_k}=Np_k(t),\\
    &\Delta^2\hat n_k=\av{\hat n_k^2}-\av{\hat n_k}^2=Np_k(t)\Big[1-p_k(t)\Big],\\
    &\sigma^2_{k,j}=-Np_k(t)p_j(t),
  \end{align}
\end{subequations}
where probabilities $p_k(t)=|\alpha_k(t)|^2$ are calculated with Eq.~\eqref{eq.sol}. The scaling of all terms from Eq.~\eqref{eq.coeffs} linearly with $N$ gives 
both $F_1$ and $F_2$ also proportional to $N$. This in turn gives the $\frac1{\sqrt N}$ dependence of the
sensitivity~\eqref{eq.sens.ml}---i.e., the shot-noise scaling with the number of atoms.

\begin{figure}[t!]
  \includegraphics[width=\columnwidth]{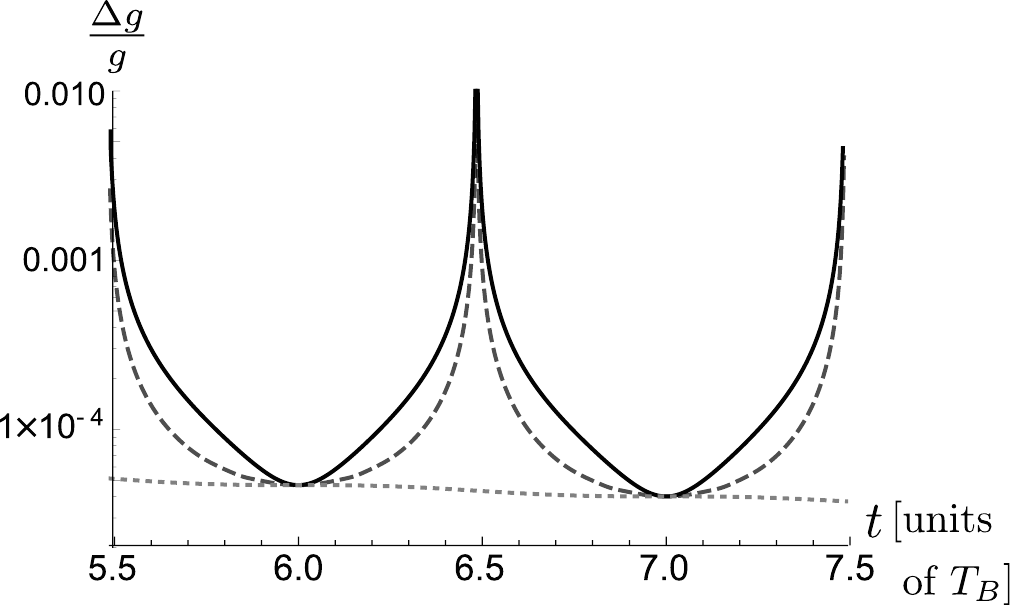}
  \caption{Relative sensitivity for $N=4\times10^4$ atoms initially loaded in a single lattice site and for 
    $J=\delta$, as a function of time. Dashed line shows the sensitivity for the measurement of the number of atoms in each site, see Eq.~\eqref{eq.sens.ml}.
    The solid black line is the error propagation for the width measurement, see Eq.~\eqref{eq.epf} with $\nu=1$. 
    The dotted line is the ultimate sensitivity calculated using the QFI, see Eq.~\eqref{eq.qfi}.}
  \label{fig.pure}
\end{figure}

The dashed line in Fig.~\ref{fig.pure} displays the sensitivity calculated with Eq.~\eqref{eq.sens.ml} using the same conditions and atom number of \ref{sec.qfi}. It is periodic
and reaches the optimal bound at the multiples of the Bloch period.
This is the first important difference between an SBOI and a standard Bloch-oscillation interferometer where the sensitivity is almost constant over the whole Bloch period. 

\subsubsection{Measurement of the width}\label{sec.width}

In this Section we compare the previous result with another method that consists in the measurement of the width of the cloud in the lattice. To this end, we introduce a (squared) width operator as
\begin{align}\label{eq.op.w}
  \hat w=\sum_k\frac{\hat n_k}Nk^2,
\end{align}
where $k=0$ is a label of the site in which initially the BEC is loaded. 
The estimation protocol consists in a measurement of the mean squared width at $M$ instants $t_1,\ldots t_M$. 
At each moment it is averaged over $\nu$ repetitions of the experiment, similarly to Eq.~\eqref{eq.mean.m}. This gives a series of averaged outcomes
\begin{align}
  \overline w_l=\frac1\nu\sum_{j=1}^\nu w_l^{(j)}\ \ \ \ \ \ \ (l=1\ldots M).
\end{align}
A theoretical curve---resulting from the averaging of the operator~\eqref{eq.op.w} over the state~\eqref{eq.sol} at time $t_l$---is fitted to this set of acquired data,  
with $g$ as a free parameter of this least-squares-fit method.  The value of $g$ obtained this way is unbiased and gives the sensitivity~\cite{chwedenczuk2010rabi}
\begin{align}\label{eq.epf}
  \Delta^2g_{\rm fit}=\frac1\nu\frac1{\sum_{l=1}^M\frac{(\av{\hat w}_l')^2}{\Delta^2\hat w_l}}.
\end{align}
From the point of view of the overall sensitivity, it is important to investigate each component of this sum, given by the error propagation formula
\begin{align}\label{eq.width.l}
  \Delta^2g_l=\frac{\Delta^2\hat w_l}{(\av{\hat w}_l')^2}.
\end{align}
The two moments are equal to
\begin{subequations}
  \begin{align}
    &\av{\hat w}_l=\sum_kp_k(t_l)k^2,\\
    &\av{\hat w^2}_l=\left(\sum_kp_k(t_l)k^2\right)^2+\frac1N\sum_kp_k(t_l)k^4.
  \end{align}
\end{subequations}
The mean $\av{\hat w}_l$ is intensive in $N$ (i.e., it does not scale with the number of atoms). This is also the case of the first part of $\av{\hat w^2}_l$, which is equal to 
$\av{\hat w}^2_l$. Therefore, in the expression for the variance, the dominant terms cancel and only the term which scales inversely with $N$ prevails, namely
\begin{align}\label{eq.var.pure}
  \Delta^2\hat w_l=\frac1N\sum_kp_k(t_l)k^4.
\end{align}
The prefactor $\frac1N$ in front of 
the sum in Eq.~\eqref{eq.var.pure} gives the shot-noise scaling of the sensitivity~\eqref{eq.epf}, as in the case of the estimation from the measurement of the number of atoms in each site 
(see Section~\ref{sec.numb}).
\begin{figure}[t!]
  \includegraphics[width=\columnwidth]{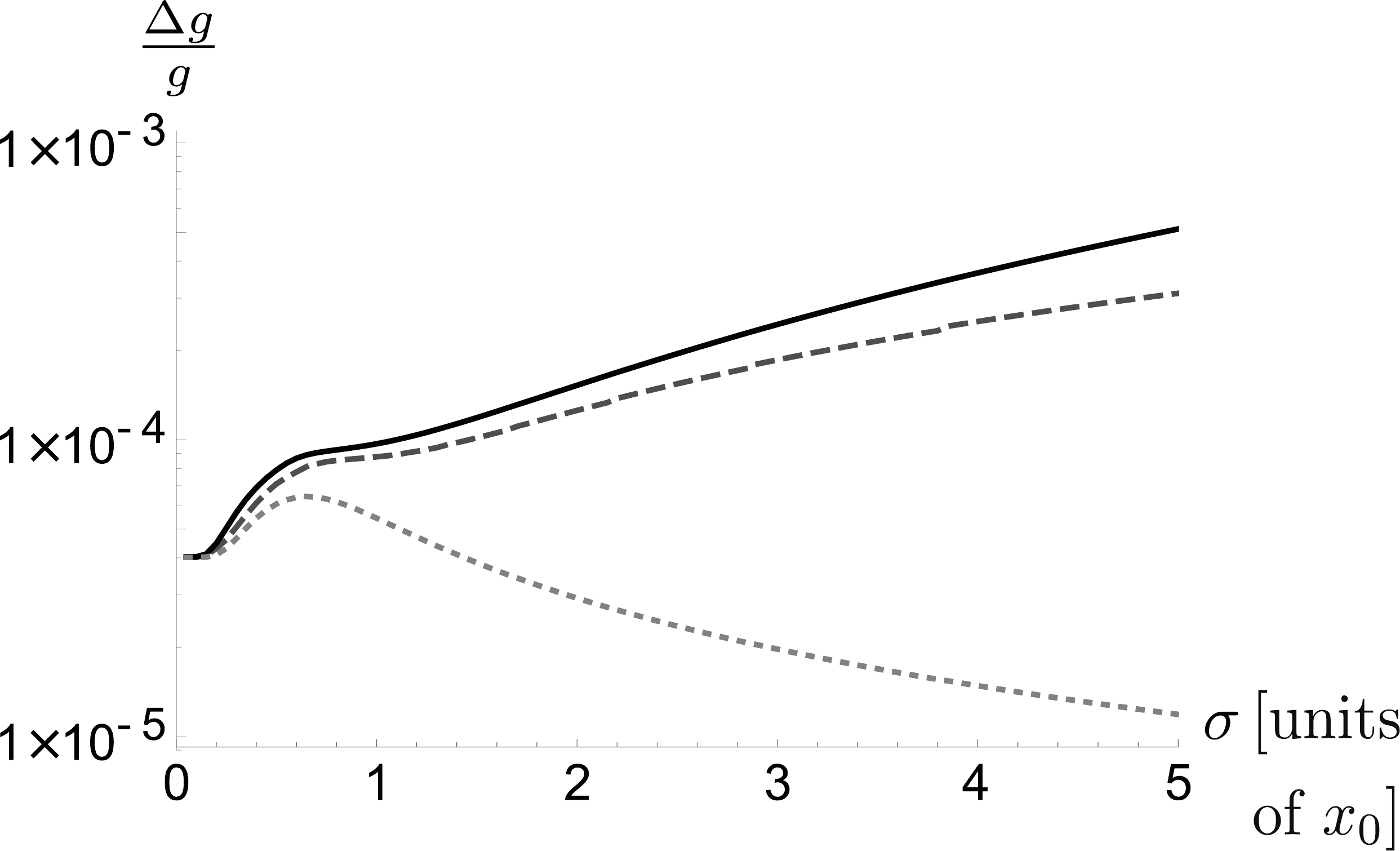}
  \caption{Sensitivities as a function of $\sigma$ (i.e., for different initial distribution of atoms in the lattice, see Eq.~\eqref{eq.init}) for $N=4\times10^4$, $t=7T_B$ and $J=\delta$.
    The ultimate bound (dotted gray), calculated with Eq.~\eqref{eq.crlb} is compared with the estimation from the number of atoms~\eqref{eq.sens.ml}---dashed black---
    and with the estimatiom from the mean width~\eqref{eq.width.l}---solid black.}
  \label{fig.init}
\end{figure}

The error propagation formula from Eq.~\eqref{eq.epf} is shown in Fig.~\ref{fig.pure} as a function of time with a solid line. Though it is worse than the sensitivity from the measurement
of the number of atoms (dashed line) it also reaches the ultimate bound at the multiples of the Bloch period. Thus we conclude that both estimation strategies discussed
in this Section can be close-to-optimal if the oscillation time is close to the Bloch period.

\subsubsection{Choice of the initial state}\label{sec.init}
So far we used a BEC localized in a single site as initial state. In this Section we investigate how the sensitivity changes 
when the atoms are initially spread over many lattice sites. For this pourpose we model the vector of coefficients $\vec\alpha(0)$ with
a Gaussian funcion
\begin{align}\label{eq.init}
  \alpha_k(0)\propto e^{-\frac{k^2}{2\sigma^2}},
\end{align}
where the proportionality sign stands for normalization. We fix $t=7T_B$ and calculate the sensitivity using the 
QFI according to Eq.~\eqref{eq.crlb} and compare it with the values predicted by Eqs~\eqref{eq.sens.ml} and~\eqref{eq.epf} as a function of the initial width of the cloud,
$\sigma_{\rm init}$. Figure~\ref{fig.init} shows the result. 

While the ultimate bound can improve as $\sigma$ increases, the sensitivities of the two estimation protocols described in Sections \ref{sec.numb} and \ref{sec.width} deteriorate. 
This means that from the point of view of these two strategies, 
the optimal operation of an SBOI requires to start with atoms loaded in a single site of the lattice. The behavior of the ultimate bound predicted by the QFI derives from the well 
known properties of a standard Bloch-oscillations interferometer where the ultimate sensitivity increases with the initial coherence length. As we will see in the next Section, an SBOI 
can recover high sensitivity operation relaxing the condition $J=\delta$ and using large values of $J$.     

\subsubsection{Dependence on the lattice parameters}\label{sec.analytic}
In order to understand how the sensitivity in Eq.~\eqref{eq.sens.ml} depends on the relevant parameters in the Hamiltonian (\ref{eq.ham}), i.e., $\delta$ and $J$, an explicit time-dependence of the on-site probability $p_k(t)$ is required. In the limit of a BEC initially localized in only one site, i.e, $p_k(0)=\delta_{k0}$, the time evolution of $p_k(t)$ is given by \cite{Hartmann}:
\begin{align}\label{eq.alpha}
  p_k(t)=\Biggr|\mathcal{J}_k\Biggr(\Bigr(\frac{4J}{\delta}\Bigr)\sin{\Bigr(\frac{\delta t}{2\hbar}-\pi n\Bigr)}\Biggr)\Biggr|^2,
\end{align}
where $\mathcal{J}_k(y)$ are Bessel functions of the first kind. We calculate the value of $\Delta\tilde g_{\rm ml}$ at optimum $t=T_B$ when both estimation strategies give the same sensitivity that
saturates the ultimate bound set by the QFI. 
At this instant the two components of the sensitivity, $F_1$ and $F_2$, have the same value equal to (see Appendix~\ref{app.a} for details)
\begin{align}\label{eq.F1.bis}
  F_1=F_2=16N\Bigr(\frac{J}{g}\Bigr)^2\Bigr(\frac{t}{\hbar}\Bigr)^2f(t),
\end{align}
where $f(t)\leqslant1$ and $f(t)=1$ for the multiples of the Bloch period. This sets the bound of the sensitivity of the estimator $\tilde g_{\rm ml}$ to the value:
\begin{align}\label{eq.sens.an}
  \Delta\tilde  g_{\rm ml}=\frac{g}{2\sqrt{2N}}\frac{1}{J}\frac{\hbar}{t}=\frac{g}{2\sqrt{2N}}\frac{1}{F \Bigr( x_0\frac{J}{\delta}\Bigr)}\frac{\hbar}{t},
\end{align} 
where $F$ is the force driving the oscillations.
If we compare this expression with the result of the numerical analysis reported in Fig. (1) for $\delta /J =1$ at $\delta t/\hbar = 7 \cdot 2\pi$ we find a perfect agreement. 

By comparing this expression with the sensitivity of a spatial Mach-Zender atom interferometer (SMZI)  the physical mechanism behind the operation of the scheme presented in this work becomes clear. 
For two modes separated by a distance $d$ in presence of an external force $F$, the accumulated phase difference $\phi = Ftd/\hbar$ detected with a shot noise $1/\sqrt{N}$ leads to an uncertainty 
\begin{align}\label{eq.sens.MZ}
  \Delta g_{\rm mzi}=\frac{\Delta \phi}{\phi} g = \frac{g}{\sqrt{N}}\frac{1}{F d}\frac{\hbar}{t}.
\end{align}  
Considering that in an SBOI the atoms, initially localized in one well, at half Bloch period reach a distance equal to the size of the Wannier Stark states, i.e. $\approx x_0 J/\delta$, 
it becomes clear---by inspecting Eq. (\ref{eq.sens.an}) and Eq. (\ref{eq.sens.MZ})---that the sensitivity of an SBOI is equal to the one of a SMZI where the separation between the two modes is of the order of 
the maximum spatial spread of the atomic wave-function over the lattice during the dynamics. 

Large separation 
between the spatial modes is crucial to have a sensitive trapped atom interferometer. This
can be easily fulfilled in an SBOI by simply increasing the tunneling energy or reducing the strength of the external force. It is the main result of our analysis.

\subsection{Experimental implementation}\label{sec.exp}

\subsubsection{Horizontal configuration}\label{sec.configuration}

Bloch-oscillation interferometers typically use a vertical optical lattice to probe the local gravitational force that corresponds to a $\delta\simeq1\mathrm{kHz} \times\hbar$ 
for lattice spacings of a fraction of a micron. In this configuration---using the maximum value of the tunneling $J$ 
that is of the order of the recoil energy $E_R=(\hbar k_L)^2/2m \approx$ few kHz $\times\hbar$---$J/\delta$ remains of the order of unity. 
The wave-function does not spread over the lattice and thus the present method cannot offer much gain with respect to the standard detection of the atomic momentum distribution in time of flight, 
for vertical lattices. 

However, in the $\delta \ll J$ limit, SBOI can be advantageous. To reach this regime, 
it could be necessary to align the optical lattice horizontally to cancel the effect of gravity and to add an external controllable force to almost compensates the one we wish to measure.  
This could limit the maximum relative sensitivity of the 
measurements due to the finite control of bias forces. However the use of an optical lattice offers the possibility to implement a controlled sweep of the phase of the lattice with 
an acceleration very close to $g$. This is a common technique used in free-falling atom interferometers where the frequencies of the Bragg or Raman lasers are chirped to remain in 
resonance with the atomic sample. In both cases the moving lattices become a reference frame respect to which the atoms feel a very small residual force.           

\subsubsection{Control of the tunneling energy}
\label{sec.tun} 

Contrary to a Bloch-oscillation interferometer, where the atomic momentum distribution does not depend on the tunneling, an SBOI requires the knowledge of $J$. 
To clarify this feature we now describe our scheme from another point of view. 

The detection of the intrap atomic density aims at identifying very precisely the Bloch period $T_B$. As described by the analysis in 
Section~\ref{sec.est} the highest sensitivity can be achieved very close to a multiple of $T_B$, where the atoms, mainly occupying the initial well, 
tunnel to the two neighbours. 
In this short time interval, using Eq.~\eqref{eq.alpha}, we find that $N_{\pm 1}:= \av{\hat n_{\pm1}}= N J^2 (t-nT_B)^2/\hbar^2$. As a consequence measuring $N$ and $N_{\pm1}$ at time $t$, 
it is possible to determine how far we are from $n T_B$ only when $J$ is known with high accuracy. Neglecting for the moment the error due to the quantum fluctuations of the atom number in the three wells, 
with a simple error propagation we find that $\Delta (t - nT_B) = (\Delta J/J) (t- nT_B)$. If the time $t$ is known precisely, then by dividing this formula with $nT_B \approx t$ we get  
\begin{align}\label{error.tunneling}
  \frac{\Delta (n T_B)}{n T_B}=\frac{\Delta g}{g}=\frac{\Delta J}{J} \frac{(t-nT_B)}{t}.
\end{align}    
This expression predicts that the relative uncertainty in the acceleration is proportional to the relative fluctuation of $J$, divided by a factor that increases the closer we perform the measurement to a multiple of the Bloch period $n T_B$. In order to confirm our simplified analysis we have performed numerical simulations as explained below.

We take into account the changes of $J$ and assume that it remains constant in each experiment but varies from shot to shot. This means that a pure state $\ket{\vec\alpha(t)}$ is replaced by a mixture
\begin{align}
  \hat\varrho=\int\!dJ\,\mathcal P(J)\,\ketbra{\vec\alpha^{(J)}(t)}{\vec\alpha^{(J)}(t)},
\end{align}
where $\mathcal P(J)$ is the probability for having $J$ and $\ket{\vec\alpha^{(J)}(t)}$ is a solution of Eq.~\eqref{eq.sch} with fixed $J$ (which appears in the Hamiltonian~\eqref{eq.ham}).
The two moments read
\begin{subequations}\label{eq.moms.mix}
  \begin{align}
    \av{\hat w}_l&=\int\! dJ\,\mathcal P(J)\sum_kp^{(J)}_k(t_l)k^2\label{eq.w.mix},\\
    \av{\hat w^2}_l&=\int\! dJ\,\mathcal P(J)\left(\sum_kp^{(J)}_k(t_l)k^2\right)^2\label{eq.w2.mix},\\
    &+\frac1N\int\! dJ\,\mathcal P(J)\sum_kp^{(J)}_k(t_l)k^4.
  \end{align}
\end{subequations}
Note that due to the fluctuations of  $J$, the dominant, intensive terms: that from line~\eqref{eq.w2.mix} and the square of the mean from line~\eqref{eq.w.mix} do not cancel, contrary to the pure-state case. 
Therefore, we expect the variance to significantly grow in presence of noise. To illustrate this effect, we take a Gaussian probability density
\begin{align}
  \mathcal P(J)=\frac1{\sqrt{2\pi}\sigma_J}e^{-\frac{(J-J_0)^2}{2\sigma_J^2}}
\end{align}
and evaluate the sensitivity using the error propagation formula from Eq.~\eqref{eq.epf} and the moments of the density operator from Eq.~\eqref{eq.moms.mix}. 

\begin{figure}[t!]
  \includegraphics[width=\columnwidth]{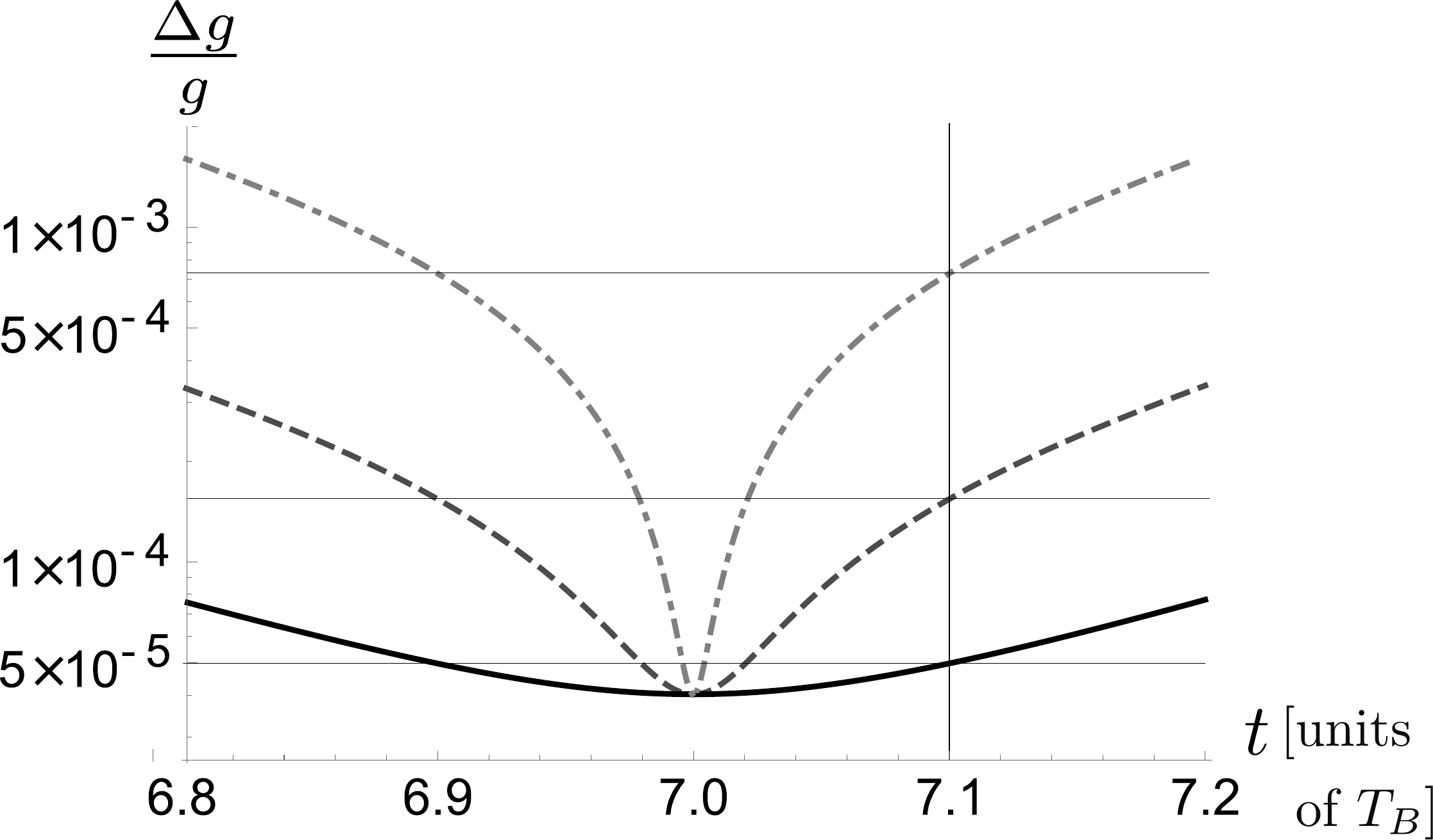}
  \caption{Sensitivity from the measurement of the width for a pure state (solid black line) compared with
    the noisy case of $\sigma_J=0.01 J_0$ (dashed dark-grey line) and $\sigma_J=0.05 J_0$ (dash-dotted light-grey line). Here, $J_0=\delta$. The vertical and 
    horizontal lines indicate the values of the sensitivity at 7.1 $T_B$ used for the comparison with the results provided by Eq.~\eqref{error.tunneling}}
  \label{fig.mixed}
\end{figure}
Figure~\ref{fig.mixed} shows the sensitivity from the width taken from Fig.~\ref{fig.pure} and compares this ideal-case result with the outcomes obtained in presence
of fluctuations of $J$ for $\sigma_J=0.01 J_0$ and $\sigma_J=0.05 J_0$, where $J_0=\delta$. The anticipated effect is clearly present, though the sensitivity remains mostly intact by the
presence of noise at the multiple of the Bloch period. We compare the numerical results at 7.1 $T_B$ for the two different levels of noise affecting $J$ reported in Fig.~\ref{fig.mixed} with the prediction provided by Eq.~\eqref{error.tunneling}. The good agreement confirms the simplified description of the interferometric scheme presented at the beginning of this section. 

\subsubsection{Finite atom number resolution}\label{sec.res}

We now incorporate finite resolution of the atom number measurement into our model. To this end, we notice that for a pure BEC, the probability for having $n_k$ atoms in $k$-th site is binomial
\begin{align}\label{eq.p1}
  p(n_k)={N\choose n_k} p_k^{n_k}(1-p_k)^{N-n_k},
\end{align}
where $p_k=|\alpha_k(t)|^2$ (contrary to before, we skip the time dependence of $p_k$, for clarity). 
The imperfection of the atom-number measurement is represented by a convolution of $p(n_k)$ with the detector resolution function $p_{\rm res}(n_k,n'_k)$ which is the probability
for obtaining $n_k$ given a true value $n_k'$. This gives
\begin{align}
  \tilde p(n_k)=\sum_{n_k'}p_{\rm res}(n_k,n'_k) p(n_k').
\end{align} 
If we approximate the probability~\eqref{eq.p1} with a normal distribution with the mean $\mu$ and the variance $\sigma$ equal to
\begin{align}
  \mu=N p_k=\av{\hat n_k},\ \ \ \sigma^2=N p_k(1-p_k)
\end{align}
we can easily include the finite atom number resolution $\sigma_{\rm res}$, taking a Gaussian $p_{\rm res}$, centered around the true value and with a width equal to the quadratic sum of $\sigma$ and $\sigma_{\rm res}$ and obtain
\begin{align}
  \tilde p(n_k)\simeq\frac{1}{\sqrt{2 \pi(\sigma_{\rm res}^2+\sigma^2)}}e^{-\frac12\frac{(n_k-\av{\hat n_k})^2}{\sigma_{res}^2+\sigma^2}}.
\end{align}
This implies that while the mean detected atom number remains unbiased, the mean square increases 
\begin{align}\label{eq.res.n2}
  \av{\hat {\tilde n}_k^2 }=\sum_{n_k=0}^N\tilde p(n_k)n_k^2\simeq \av{\hat n_k^2 }+\sigma_{\rm res}^2.
\end{align}
Also, since in this model of finite resolution, the probabilities $p_{\rm res}$ at different sites $k$ and $k'\neq k$ are independent, the average 
$\av{\hat n_k\hat n_{k'} }$ is unaltered. 
\begin{figure}[t!]
  \includegraphics[width=\columnwidth]{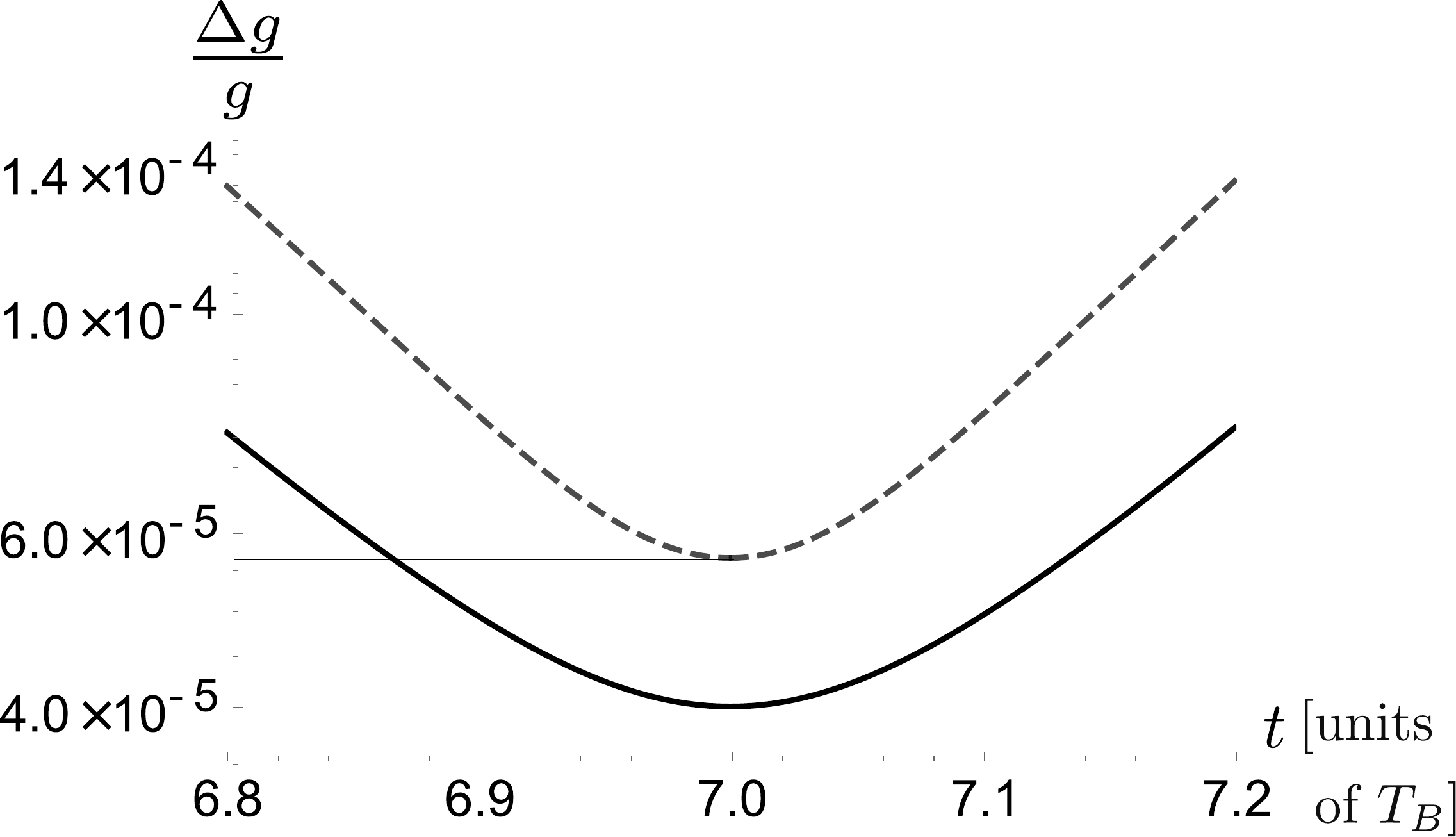}
  \caption{The effect of finite resolution on the sensitivity from the measurement of the width. Solid black line is the ideal case $\lambda=0$, 
    and the dashed black line is for $\lambda=1$.}
  \label{fig.res}
\end{figure}

We now use these results to calculate the impact of resolution on the sensitivity from Eq.~\eqref{eq.epf}. The mean of $\hat w$ remains unchanged but the variance increases since according to
Eq.~\eqref{eq.op.w}
\begin{align}
  \av{\hat w^2}=\sum_{k\neq k'}\frac{\av{\hat n_k\hat n_{k'}}}{N^2}k^2k'^2+\sum_k\frac{\av{\hat n_k^2}}{N^2}k^4.
\end{align}
While the first ``off-diagonal'' part is intact by the finite resolution, the second ``diagonal'' term is modified according to Eq.~\eqref{eq.res.n2}. 

Fig. ~\ref{fig.res} shows the impact of
this imperfection on the sensitivity, assuming that at each site, the detector's resolution is proportional to the shot-noise fluctuations of the mean atom number at this site, i.e., 
$\sigma_{\rm res}^2=\lambda\av{\hat n_k}$ ($\lambda$ is the proportionality constant). 
The minimal $\Delta g$ at $t=7T_B$ increases from the limit set by the QFI $\Delta g=4.019\times10^{-5}g$ for 
$\lambda =0$, to $5.684\times10^{-5}g$ for $\lambda=1$, i.e. a factor $\sqrt{2}$. This is expected considering that for $\lambda =1$ the detection noise is equal to the shot noise atom number fluctuation for each site. 

An approximate analytical formula to quantify the effect of a finite atom number resolution on the sensitivity can be again derived at times close to a multiple of the Bloch period using the formula $N_{\pm 1}=NJ^2(t-nT_B)^2/\hbar^2$ and the error propagation of $\Delta N_{\pm 1}$. 
To confirm the validity of this approach we use it to derive the sensitivity bound assuming the shot noise scaling $\Delta N_{\pm 1} =\sqrt{N_{\pm 1}}$. 
For negligible fluctuations of the tunneling energy we get $\Delta (t-nT_B)/(t-nT_B)=\Delta N_{\pm}/(2N_{\pm})=1/(2\sqrt{N_{\pm}})=\hbar/[ 2\sqrt{N}J(t-nT_B)]$. It follows that 
\begin{align}\label{error.tunneling2}
  \frac{\Delta (n T_B)}{n T_B}=\frac{\Delta g}{g}=\frac{\hbar}{2\sqrt{2N}J t},
\end{align} 
where the additional $\sqrt{2}$ takes into account the double measurement on the two neighboring sites. Note the perfect agreement between this formula and Eq.~\eqref{eq.sens.an}
derived with a rigorous calculation.

\begin{figure}[t!]
  \includegraphics[width=\columnwidth]{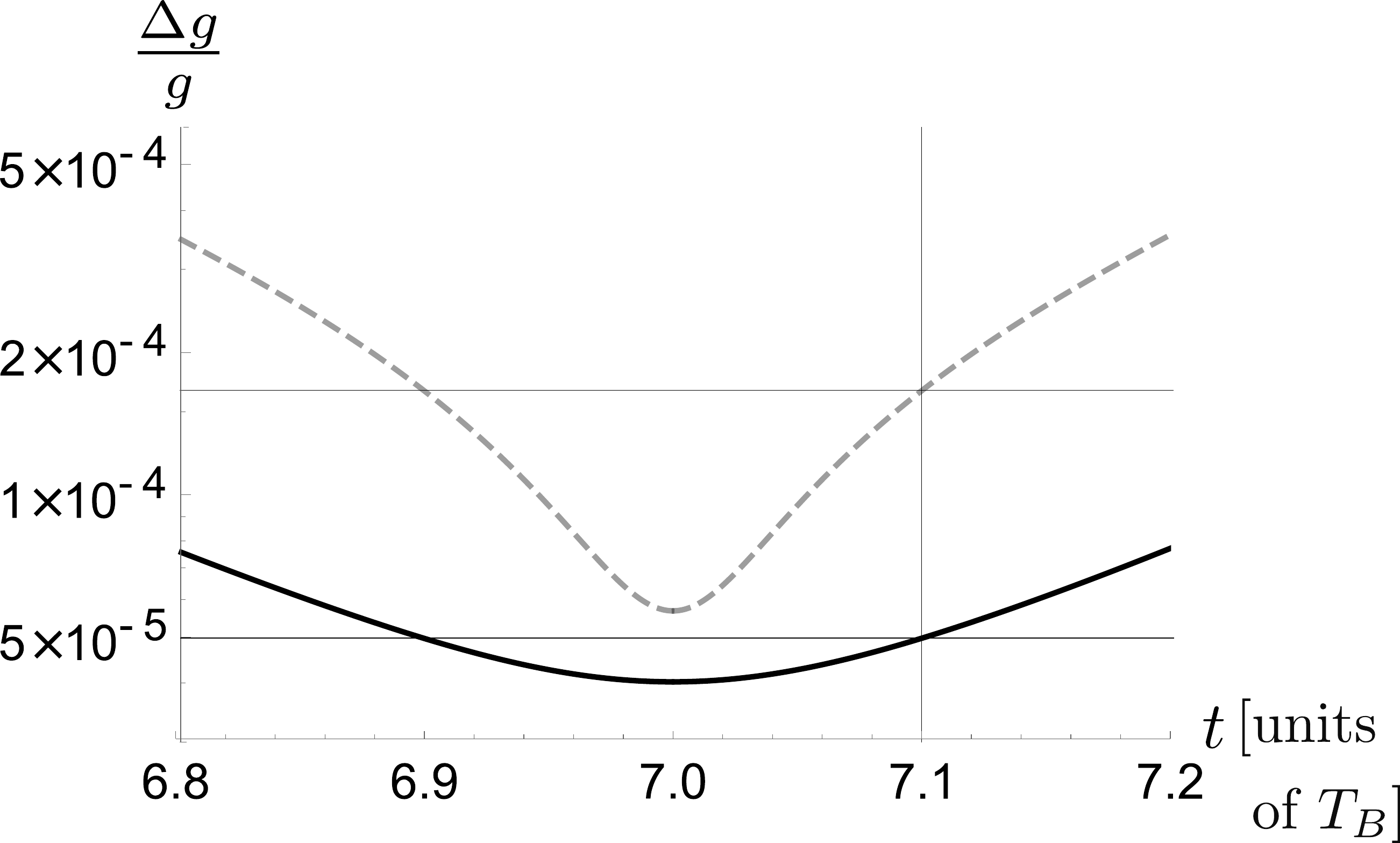}
  \caption{The combined effect of the fluctuations of the tunneling constant ($\sigma_J=0.01$) and finite resolution ($\lambda=1$) on the sensitivity from the measurement of the width (dahsed black).
    Solid black line is the ideal no-noise case.}
  \label{fig.comb}
\end{figure}

Finally, in Fig.~\ref{fig.comb} we show the combined effect of fluctuations of the tunneling constant and finite resolution 
using $\sigma_J=0.01\delta$ and $\lambda=1$. While the sensitivity at the optimal time does not shfit from the $\sigma_J=0$ case ($\Delta g=5.684\times10^{-5}g$), the non-zero $\sigma_J$ casuses
the region where the width measurement is close-to-optimal to shrink with respect to Fig.~\ref{fig.res}, similarly to the effect observed in Fig.~\ref{fig.mixed}.

\subsubsection{Lattice spacing}\label{sec.lat_spec}
In this paragraph we identify the optimal value of the lattice spacing. 
For a standard Bloch-oscillation interferometer the sensitivity does not depend on $x_0$ but can be enhanced only reducing the initial width of the atomic momentum distribution.
Also for an SBOI the sensitivity~\eqref{eq.sens.an} apparently does not depend on $x_0$. However some experimental constraints change the overall picture. 
The use of an optical lattice with $x_0$ equal to a fraction of a micron makes impossible to load many atoms in a single lattice site due to high three body losses when atomic densities are big. 
This limits the improvement from the shot-noise scaling, i.e., with the inverse of $\sqrt{N}$. 

In addition, if the lattice spacing is too small, it is  challenging to precisely count the atoms in each site. As a consequence larger lattice spacing naturally improves the sensitivity of an SBOI. 
To determine the optimal lattice spacing we notice that the relative uncertainty in an SBOI is limited by the relative fluctuations of $J$ and the atomic shot-noise. As a consequence,
high sensitivity can be achieved compensating the external force with an accurate bias field and operating the interferometer with a very small residual force $mg$. 

However, the sensitivity saturates the optimal bound (set by the $F_q$) at multiples of the Bloch period. Considering that in real experiments the interrogation time $\tau$ is finite, 
due to decoherence induced by residual interactions or experimental noise, we cannot work with arbitrarily small forces but keep $T_B =h/(mgx_0) < \tau$. 
This condition suggests to increase $x_0$ while reducing $g$. However, an SBOI requires to keep the tunneling $J$ sufficiently high to spread the wave-function over few lattice sites, i.e., 
$J \geqslant \delta = mgx_0$. Using the maximal value of $J$ as a function of $x_0$, i.e. $J=\hbar^2 \pi^2/(8mx_0^2)$ we obtain an upper bound on $x_0$, which
sets the minimal applicable acceleration as a function of the coherence time $\tau$          
\begin{align}\label{eq.gmin}
  g_{\rm min}=\frac{8}{\tau^{3/2}}\sqrt{\frac{\pi \hbar}{m}}
\end{align} 
and the required spacing is $x_0 = \sqrt{\pi \hbar \tau /m}/4$.

Finally, we discuss how the lattice spacing influences the control on the tunneling energy. 
As indicated in Eq.~\eqref{eq.sens.MZ}, 
the sensitivity depends on the spread of the wave-function during the dynamics that is equal to $x_0 J/\delta = J/(mg)$. 
As a consequence, it is directly related to the spatial resolution of the sensor. If we consider applications where this quantity is determined by the measurement constraints, 
fixing $g$ corresponds to fixing $J$. In the tight binding approximation, the tunneling energy $J$ in unit of $E_R$ depends on the lattice depth $s_L$ through  a scaling factor $s_L^{3/4}e^{-2\sqrt{s_L}}$. 
It is possible to demonstrate via a simple error propagation that the relative fluctuation of the tunneling constant, that directly affects the sensitivity as shown in Sec. \ref{sec.tun}, 
depends on the relative fluctuation of the lattice depth $\Delta s_L/s_L$ by the relation $\Delta J/J\sim\sqrt{s_L}\Delta s_L/s_L$. 
This expression shows that, the larger the value of $s_L$ needed to achieve a specific value of $J$, the larger the constant of proportionality. 
Therefore, bigger $x_0$ reduces the fluctuations of $J$ provided a specific instability of the lattice depth. Similar argument is valid also in the limit of small lattice depths.   

\subsubsection{Final remarks}\label{sec.example}
In this last section we consider a realistic example of an SBOI.
We take  $\tau\sim1$ s and $N\sim10^{4}$, and from Eq.~\eqref{eq.gmin} we get the smallest measurable acceleration $\sim 5\times10^{-5}g$ and an optimal lattice spacing of $x_0=17 \mu$m. 
From Eq.~\eqref{eq.sens.an}, if we neglect fluctuations of the tunneling energy, the relative uncertainty is $4\cdot10^{-4}$ and a single shot ($\nu=1$) sensitivity is of the order of $10^{-8}g$. 
With an improvement of a factor of ten in the coherence time, it is possible to reach a sensitivity  comparable with the state-of-the-art but with an unprecedented 
spatial resolution of the order of 100 $\mu$m. 
Note that in order to achieve comparable sensitivities with a standard Bloch-oscillation interferometer, the sensor should be operated with an ideal BEC with an initial coherence length of 100 $\mu$m 
that is not within the reach of current ultra-cold atoms technology. 

The advantage of the setup discussed in this work with respect to a standard Bloch-oscillation interferometer
is that the sensitivity depends on the amplitude of the oscillations of the cloud in the lattice,
rathern than on a large initial extension of the condensate.

The main obstacles to the operation of an SBOI 
is the reduction of the atom interactions~\cite{PhysRevLett.100.080404, fattori, PhysRevA.86.033421} and the realization of optical lattices with large spacings. 
Carbon dioxide gas lasers can be used to generate optical lattices with sites separation of $\sim 5$ microns \cite{PhysRevA.62.051801}. 
In the near future mid-infrared cw high power radiation generated with quantum cascade lasers might broaden the spectrum of available spacings. 
Arbitrarily large separations between lattice sites could be finally achieved using recently realized Beat-note optical lattices \cite{Masi}.  

\section{Conclusions and acknowledgements}\label{sec.ack}

In this work, we studied a matter-wave interferometer consisting of a BEC undergoing Bloch oscillations in an optical lattice. 
We assumed that the parameter---here the acceleration $g$---is estimated from the in-situ measurement of the atomic density. 
We considered the case, when the lattice is oriented almost horizontally, so that the increment of the linear potential from site to site is smaller than the tunneling energy.
In this regime, atoms spread over many sites, so the wave-function probes the perturbing potential over a large distance.

Using the metrological tool known as the quantum Fisher information, we have calculated the best-achievable sensitivity $\Delta g$ for this configuration, 
and showed that indeed the precision benefits from the large extension of the cloud. Having established the ultimate bound, we have determined the sensitivity for two experimental scenarios:
when $g$ is estimated from the measurement of the number of atoms in each site or from the width of the atomic cloud. As the latter carries less information with respect to the former, 
it gives an inferior sensitivity, apart from the vicinity of the multiples of the Bloch period. At these times, all three sensitivities (i.e., obtained from the QFI and the two protocols), coincide.

We incorporated two sources of imperfections: fluctuations of the tunneling constant and the limited resolution of the atom-number measurement. With both these deficiencies, 
the sensitivity drops but remains competitive to results obtained with the state-of-the-art settings. 
We conculde by stating that the matter-wave interferometer proposed  here turns out to be a promising solution for compact sensors or for the measurements of small forces with high spatial resolution.

IN and JC are supported by Project no. 2017/25/Z/ST2/03039, funded by the National Science Centre, Poland, under the QuantERA programme. This work was supported by the project TAIOL of QuantERA ERA-NET Cofund in Quantum Technologies (Grant Agreement No. 731473) implemented within the European Union’s Horizon 2020 Programme.

\appendix

\section{Analytic expression of the sensitivity}\label{app.a}

Here, we present the detailed derivation of Eqs~\eqref{eq.F1.bis} and~\eqref{eq.sens.an}.
We start with the $F_1$, given by Eq.~\eqref{eq.f1}. Using the expression for $p_k(t)$ from Eq.~\eqref{eq.alpha} and the moments of the atom-number operator from Eq.~\eqref{eq.coeffs}, we obtain
\begin{align}
  \av{\hat{n}_{k}}'=N|\mathcal{J}_k(y)|(\mathcal{J}_{k-1}(y)-\mathcal{J}_{k+1}(y))y',
\end{align}
where we used $\mathcal{J}'_k(y)=1/2(\mathcal{J}_{k-1}(y)-\mathcal{J}_{k+1}(y))$ and introduced a function of $\delta$
\begin{align}
  y=\frac{4J}{\delta}\sin{\Bigr(\frac{\delta t}{2\hbar}-\pi n\Bigr)}.
\end{align}
Its derivative is equal to
\begin{align}
  y'=\frac{4J}{\delta^2}\Biggr[-\sin{\Bigr(\frac{\delta t}{2\hbar}-\pi n\Bigr)}+\frac{\delta t}{2\hbar}\cos{\Bigr(\frac{\delta t}{2\hbar}-\pi n\Bigr)}\Biggr].
\end{align}
When the measurement is performed  after many Bloch periods,  the second term in the parenthesis dominates leading to an approximate expression
\begin{align}
  \av{\hat{n}_{k}}'&\simeq|\mathcal{J}_k(y)|(\mathcal{J}_{k-1}(y)-\mathcal{J}_{k+1}(y))\nonumber\\
  &\times\frac{2J}{\delta}\frac{t}{\hbar}\cos{\Bigr(\frac{\delta t}{2\hbar}-\pi n\Bigr)}.\label{partial_K}
\end{align}
The variance of the atom-number operator is simply
\begin{align}\label{delta_K}
  \Delta^2\hat{n}_k=N|\mathcal{J}_k(y)|^2(1-|\mathcal{J}_k(y)|^2).
\end{align}
Bringing together Eqs~\eqref{partial_K} and~\eqref{delta_K} gives
\begin{align}\label{eq.app.f1}
  F_1=\sum_k\frac{(\av{\hat n_k}')^2}{\Delta^2\hat n_k}=16\Bigr(\frac{J}{\delta}\Bigr)^2\Bigr( \frac{t}{\hbar}\Bigr)^2f(t),
\end{align}
where $f(t)$ is the time-dependent function
\begin{align}
  f(t)=\frac{1}{4}\cos^2{\Bigr(\frac{\delta t}{2\hbar}-\pi n\Bigr)}\sum_{k=1}^M\frac{|\mathcal{J}_{k-1}(y)-\mathcal{J}_{k+1}(y)|^2}{1-|\mathcal{J}_k(y)|^2}
  \label{Fisher_fin}
\end{align}
reaches its maximum $f(t)=1$ at the multiples of the Bloch period, $t=nT_B$, $n\in\mathbb N$. 

In the next step, we calculate the $F_1$ and $F_2$ in the vicinity of $t=nT_B$, when almost all atoms are located in the central site and only a small fraction is present in the two neighboring sites. 
Therefore
\begin{align}\label{eq.eps}
  p_{\pm1}=\epsilon,\ \ \ p_0=1-2\epsilon,\ \ \ (\epsilon\ll1).
\end{align}
The approximate expression for the $F_1$, obtained from Eq.~\eqref{eq.f1} gives
\begin{align}
  F_1\simeq N\frac{(p_0')^2}{p_0(1-p_0)}+2N\frac{(p_1')^2}{p_1(1-p_1)},
\end{align}
where we used the symmetry between $\pm1$ (hence the factor of 2) and $\simeq$ stands for the dropping of $p_{\pm 2}$, {\it etc}. Plugging Eq.~\eqref{eq.eps} above, we obtain
\begin{align}
  F_1\simeq N\frac{(1-2\epsilon)'^2}{(1-2\epsilon)2\epsilon}+2N\frac{(\epsilon')^2}{\epsilon(1-\epsilon)}\simeq4N\frac{\epsilon'^2}{\epsilon}.
\end{align}
We now calculate the $F_2$. Using the expression for the moments of the atom number operator, we have directly from Eq.~\eqref{eq.f2}
\begin{widetext}
  \begin{align}
    F_2&=N\sum_{k\neq j}\frac{p_k'}{p_k(1-p_k)}\frac{p_j'}{p_j(1-p_j)}(-p_kp_j)=-N\sum_{k\neq j}\frac{p_k'}{(1-p_k)}\frac{p_j'}{(1-p_j)}=\nonumber\\
    &-N\sum_{k,j}\frac{p_k'}{(1-p_k)}\frac{p_j'}{(1-p_j)}+N\sum_{k=j}\frac{p_k'}{(1-p_k)}\frac{p_j'}{(1-p_j)}=
    -N\left(\sum_{k}\frac{p_k'}{(1-p_k)}\right)^2+N\sum_{k}\left(\frac{p_k'}{(1-p_k)}\right)^2=\nonumber\\
    &=-N\left(\frac{-2\epsilon'}{2\epsilon}+2\frac{\epsilon'}{1-\epsilon}\right)^2+N\Bigg[\left(\frac{-2\epsilon'}{2\epsilon}\right)^2+2\left(\frac{\epsilon'}{1-\epsilon}\right)^2\Bigg]
    =4N\frac{\epsilon'^2}{\epsilon}-2N\left(\frac{\epsilon'}{1-\epsilon}\right)^2\simeq4N\frac{\epsilon'^2}{\epsilon}=F_1.
  \end{align}
\end{widetext}
Thus we showed that when $\epsilon\rightarrow0$ (thus $t\rightarrow T_B$), the $F_1$ and $F_2$ are equal. Therefore, using formula for the sensitivity~\eqref{eq.sens.ml}, which for a single shot ($\nu=1$) is
\begin{align}
  \Delta^{2}\tilde g_{\rm ml}=\frac{F_1+F_2}{F_1^2}, 
\end{align} 
we obtain
\begin{align}\label{eq.delta.fin}
  \Delta^{2}\tilde g_{\rm ml}=\frac{2}{F_1}.
\end{align} 
This results, combined with Eq.~\eqref{eq.app.f1}, gives the Eq.~\eqref{eq.sens.an} from the main text.

\end{document}